\documentclass[useAMS,usenatbib]{mn2e}

\title[Magnetic field in Cepheus A]
{Magnetic field in Cepheus A as deduced from OH maser polarimetric observations} 
\author[A.~Bartkiewicz et al.]
       {A.~Bartkiewicz,$^1$\thanks{formerly Niezurawska, e-mail:
annan@astro.uni.torun.pl} M.~Szymczak,$^1$
        R.J.~Cohen$^2$ and A.M.S.~Richards$^2$ \\
       $^1$Toru\'n Centre for Astronomy, Nicolaus Copernicus 
          University, ul.Gagarina 11, 87-100 Toru\'n, Poland \\
       $^2$Jodrell Bank Observatory, University of Manchester,
          Macclesfield, Cheshire SK11 9DL, UK}

\date{Released 2004 Xxxxx XX}

\pagerange{\pageref{firstpage}--\pageref{lastpage}} \pubyear{2004}

\def\LaTeX{L\kern-.36em\raise.3ex\hbox{a}\kern-.15em
    T\kern-.1667em\lower.7ex\hbox{E}\kern-.125emX}

\usepackage{graphicx}
\begin{document}

\label{firstpage}

\maketitle

\begin{abstract}
We present the results of MERLIN polarization mapping of OH masers at
1665 and 1667\,MHz towards the Cepheus\,A star-forming region.  The
maser emission is spread over a region of 6 arcsec by 10 arcsec,
twice the extent previously detected.  In contrast to the 22-GHz water
masers, the OH masers associated with H{\small II} regions show
neither clear velocity gradients nor regular structures.  We
identified ten Zeeman pairs which imply a magnetic field strength
along the line-of-sight from $-$17.3 to $+$12.7\,mG. The magnetic
field is organised on the arcsecond scale, pointing towards us in the
west and away from us in the east side.  The linearly polarized
components, detected for the first time, show regularities in the
polarization position angles depending on their position. The electric
vectors of OH masers observed towards the outer parts of H{\small II}
regions are consistent with the interstellar magnetic field
orientation, while those seen towards the centres of H{\small II}
regions are parallel to the radio-jets. 
A Zeeman quartet inside a southern H{\small II} region has now been
monitored for 25 years; we confirm that the magnetic field decays
monotonically over that period.  

\end{abstract}
\begin{keywords}
masers $-$ polarization $-$ magnetic fields $-$ stars: formation $-$
ISM: individual: Cep~A $-$ molecules $-$ radio lines: ISM $-$ H{\small
II} regions
\end{keywords}

\section{Introduction}
The connection between interstellar magnetic fields and 
the phenomenon of bipolar outflow from young stars is a topic of great 
observational and theoretical interest (e.g.\,\citealt{bachiller96}; 
\citealt{ward-thompson00}; \citealt{kudoh97};  and references 
therein).  OH masers provide a useful tool for probing the kinematics 
and the magnetic field on subarcsecond scales, near the centres of 
bipolar outflows from massive young stars \citep{cohen89}.  The OH masers 
are generally found at distances from a few hundred to a few thousand 
au from the central source, and are sensitive to magnetic fields as 
small as 1~mG through 
the Zeeman effect.  A series of OH maser studies has revealed a common 
pattern of OH masers tracing a dense molecular disc perpendicular 
to the bipolar outflow, with the magnetic field twisted in the way 
predicted by the magnetohydromagnetic models by \citet{uchida85} 
and others (Hutawarakorn \& Cohen 1999, 2003, 2005;
\citealt*{hutawarakorn02}; \citealt*{gray03}). The magnetic field 
is oriented along the outflow direction on the 10-arcsec scales probed 
by submm-polarimetry, but has a toroidal component wound around the 
outflow on the arcsec scale probed by the OH masers.  This appears in 
the form of a magnetic field reversal on opposite sides of the disc.  
In the present 
paper we consider the OH masers associated with the well known 
outflow source Cepheus\,A.  

Cepheus\,A is a nearby star-forming region (within 1\,kpc)
and the densest part of the molecular cloud complex located
to the south of the Cepheus OB3 association (\citealt{sargent77}).  The
large-scale ($\sim$10 arcmin) morphology of this cloud seen in CO
emission implies bipolar outflow at velocities of
$\sim$25\,km\,s$^{-1}$ (\citealt*{rodriguez80}).  The base region of
Cep\,A observed in the NH$_3$ and CO lines with angular resolutions of
2 arcsec and 15 arcsec respectively, shows a quadrupolar structure of
the molecular outflow (\citealt{torrelles93}). The observations of
several thermal lines confirm the presence of multiple outflows
(\citealt{narayanan96}; \citealt{codella03}). The complexity of the
spatial and velocity structure of the Cep\,A region, characteristic of
turbulent flow, is supported by maps of the H$_2$ emission line at
2.1$\mu$m (\citealt*{hiriart04}).  \citet{hughes84} found several
compact H{\small II} regions (CHIIRs) at centimetre wavelengths; these
mostly follow the edges of the NH$_3$ clouds
(\citealt{torrelles93}). All of those CHIIRs are younger than
10$^3$\,yr and some of them are likely associated with high-mass
powering stars of spectral types B3 or earlier
(\citealt{hughes01}). Some CHIIRs are accompanied by clusters of OH,
H$_2$O and CH$_3$OH masers (e.g.\/ \citealt*{cohen84};
\citealt{torrelles96}; \citealt*{minier00}). Systematic studies of OH
masers with high angular resolution revealed that most OH components
move away from the central CHIIRs on an estimated outflow time-scale
of $\sim$300\,yr (\citealt*{migenes92}).  Observations of H$_2$O
masers indicate the presence of remarkable linear and arcuate
structures, which can delineate shock fronts or discs
(\citealt{torrelles98}; \citealt{torrelles01}; \citealt{curiel02};
\citealt{gallimore03}). The H$_2$O maser structure centred on the
source HW2 is interpreted as due to a molecular disc, 600\,au in
diameter, which is perpendicular to the thermal radio jet. The inferred
mass of the powering source is $\sim$20~M$_\odot$, i.e.\/ it is a B0.5 or
earlier star (\citealt{torrelles96}).

In the paper we report full polarimetric observations of Cep\,A in the
1.6\,GHz OH maser lines. 
Zeeman splitting of the OH maser lines at 1665- and 1667\,MHz gave a value
of $+$3.5 mG, where the plus sign indicates a field directed away from us.
Another Zeeman pair has been identified in the 22-GHz water maser line;
the inferred field strength is $-$3.2 mG \citep{sarma02}, that is,
pointing towards us. Such magnetic fields are believed to have
an non-negligible dynamical effect \citep{migenes92}.  We attempt to
extend previous the observations by using new polarization data to
obtain an overall structure of magnetic field in the central region of
Cep\,A.

\begin{table*}
\caption {Details of the MERLIN observations.}
\begin{tabular}{cccccrcc}
\hline
      &            &\multicolumn{2}{c}{ Phase calibrator}&\multicolumn{2}{c}{ Synthesised beam} & & \\
 Date & Line transition & Frequency & Flux density   & HPBW  & \multicolumn{1}{c}{P.A.} & Time  & Rms noise \\
      &            &            &                    &       &     &  on source & per channel \\
      &  (MHz)     &  (MHz)     & (mJy )  & (mas)& \multicolumn{1}{c}{(\degr)}    & (h) & (mJy\,b$^{-1}$) \\
\hline
20 May 1999 & 1612.231 & 1612.4 & 124 & 180$\times$140 & $-$27 & 7.5 & 3 \\
            & 1665.402 & 1661.2 & 128 & 149$\times$120 & $-$7  & 6.4 & 5 \\
            & 1667.359 & 1661.2 & 125 & 148$\times$119 & $-$11 & 6.4 & 4 \\
            & 1720.530 & 1720.8 & 130 & 154$\times$107 & $-$24 & 7.6 & 3 \\
30 May 1999 & 1665.402 & 1661.2 & 128 & 136$\times$123 & 2     & 5.2 & 3 \\
            & 1667.359 & 1661.2 & 130 & 136$\times$122 & $-$5  & 5.1 & 4 \\
16 June 1999& 1612.231 & 1612.4 & 124 & 165$\times$115 & $-$33 & 3.7 & 4 \\
            & 1720.530 & 1720.8 & 150 & 143$\times$109 & $-$32 & 3.7 & 4 \\                              
\hline
\multicolumn{8}{l}{Antennas: Defford, Cambridge, Knockin, Darnhall, Mark\,II,
Tabley} \\
\end{tabular}
\end{table*}

\section{Observations and data reduction} 
Cep\,A was observed in 1999 in all four ground-state OH transitions
using 6 telescopes of the Multi-Element Radio Linked Interferometer Network
(MERLIN) (Table 1).  Each transition was observed
in left- (LHC) and right-circular polarization (RHC) simultaneously
and correlated to obtain all four Stokes parameters.  The spectral
bandwidth of 250~kHz, covering a velocity range of 45\,km\,s$^{-1}$,
was divided into 128 channels, yielding a channel spacing of
0.35\,km\,s$^{-1}$. The band centre was set to a local standard of
rest (LSR) velocity of $-$14\,km\,s$^{-1}$.

The phase-referencing technique was applied at all epochs with 
the continuum source 2300+638 used as a phase-calibrator.
Since the reference source was too weak 
to be observed in narrow-band mode, it was observed in wide-band
mode (16-MHz) at the appropriate frequencies  (given in Table 1)
to achieve the optimum signal-to-noise ratio.
The cycle-times between Cep\,A and the reference were 5.5\,min+2\,min at the
main line transitions and 6.5\,min+2\,min at the satellite line transitions.
In addition 3C286 and 3C84 were observed for 1-3 hours in all sessions, at 
the relevant narrow and wide-band frequencies corresponding to the 
frequencies given in Table 1 (corrected to the appropriate $V_{\rm LSR}$ 
for the narrow-band configurations). 3C286 was used as the primary calibrator for
the flux density scale and for the polarization angle;
3C84 was used as the point source, bandpass calibrator and for the polarization
leakage corrections.

Data reduction was carried out using standard procedures
\citep{diamond03} with the local {\tt d-programs} at Jodrell Bank and
the Astronomical Image Processing System ({\sc aips}).
To map the emission we used a circular Gaussian beam of full width half 
maximum (FWHM) 120\,mas and the pixel separation was 40\,mas in all epochs
of observations (the synthesised beams for all data are given in Table 1). 
The  rms noise ($\sigma$) 
levels in emission-free Stokes $I$ single channel maps were typically a  
few mJy\,beam$^{-1}$ (Table 1). The 1665-MHz data taken on May 20 were 
affected by the ringing effect. This data set was therefore smoothed 
using a Hanning function.

A region of 20 arcsec by 20 arcsec centred at RA$=22^{\rm
h}56^{\rm m}18\fs033$, Dec$=62^o 01' 48\farcs347$ (J2000) was inspected for
total intensity emission ($I$ Stokes parameter) 
stronger than 5$\sigma$.  We fitted 2 dimensional
Gaussian components to each patch of maser emission in each channel to measure
its position and flux density. The emission was considered to be real if
present at least in three contiguous channels. We measured the flux
densities in
the images of the other Stokes parameters at the position of each $I$
Stokes component, using the local {\sc aips} task {\sc mfquv}. 
We grouped series of components above $5\sigma$ arising from similar positions 
in successive channels into features and analysed them. 
More details on data reduction
procedures and determination of the uncertainties in the maser
positions were given by \citet{niezurawska04}.

\section{Results}
\subsection{Polarization spectra}

\begin{figure}
\resizebox{\hsize}{!}{\includegraphics{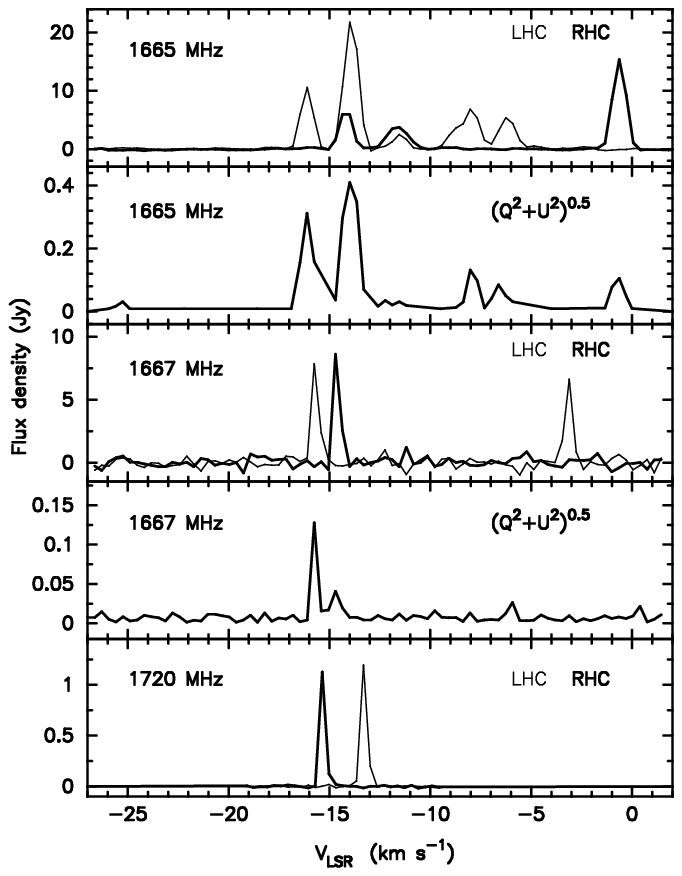}}
\caption{The spectra of Cep~A taken with MERLIN in May 1999. OH main-line
spectra are averaged from two epochs (20 and 30 May). The bottom panel shows
the 1720-MHz emission detected in 20 May 1999.}
\label{fig1} 
\end{figure}

\begin{table*}
\caption {Positions of OH masers in Cep\,A observed on 1999 May 30. The newly
detected components are marked with superscript $n$. The numbers in brackets indicate
relative errors in positions ($\times$0\fs0001 for RA, $\times$0\farcs001
for Dec).}
\begin{tabular}{lcccc|lcccc}
\hline
V$_{\rm LSR}$ & RA   & Dec & S$_{\rm p}$ & Feature & V$_{\rm LSR}$ & RA   & Dec &
S$_{\rm p}$ & Feature \\
(km\,s$^{-1}$) & (22$^{\rm h}$56$^{\rm m}$) & (62\degr01\arcmin) 
& (Jy\,b$^{-1}$) & 
& (km\,s$^{-1}$) & (22$^{\rm h}$56$^{\rm m}$) & (62\degr01\arcmin) &
(Jy\,b$^{-1}$) & \\
\hline
\multicolumn{5}{|l|}{1665-MHz LHC} & \multicolumn{5}{|l|}{1665-MHz RHC} \\
$-$25.2$^n$ & 18$\fs$1642(2) & 49$\farcs$668(2) &  0.285 & A & $-$18.0$^n$ & 18$\fs$1646(3) & 49$\farcs$673(2) & 0.174 & a \\
$-$16.1$^n$ & 18$\fs$0771(2) & 45$\farcs$809(8) &  0.296 & B & $-$14.2     & 18$\fs$1143(1) & 46$\farcs$316(1) & 5.240 & b \\
$-$16.1$^n$ & 17$\fs$9472(3) & 49$\farcs$829(3) &  0.779 & C & $-$14.2$^n$ & 17$\fs$9474(4) & 49$\farcs$839(3) & 0.359 & c \\
$-$16.1     & 18$\fs$1148(1) & 46$\farcs$320(1) &  8.880 & D & $-$13.1     & 18$\fs$0976(2) & 49$\farcs$343(2) & 0.359 & d \\
$-$14.4$^n$ & 18$\fs$1383(2) & 46$\farcs$466(1) &  3.335 & E & $-$12.3$^n$ & 18$\fs$1464(3) & 47$\farcs$030(3) & 0.301 & e \\
$-$13.9$^n$ & 18$\fs$1324(1) & 46$\farcs$494(1) & 16.250 & F & $-$11.9$^n$ & 18$\fs$2134(15)& 46$\farcs$510(13)& 0.066 & f \\
$-$13.9$^n$ & 18$\fs$0951(8) & 45$\farcs$964(6) &  0.579 & G & $-$11.7$^n$ & 17$\fs$9699(6) & 50$\farcs$012(5) & 0.130 & g \\
$-$13.9     & 17$\fs$9651(3) & 50$\farcs$006(2) &  1.426 & H & $-$11.6$^n$ & 18$\fs$1352(1) & 46$\farcs$487(1) & 2.070 & h \\
$-$12.3     & 17$\fs$8382(2) & 48$\farcs$675(1) &  0.622 & I & $-$11.5     & 17$\fs$8662(5) & 48$\farcs$812(4) & 0.196 & i \\
$-$11.5     & 17$\fs$8656(2) & 48$\farcs$813(1) &  2.224 & J & $-$11.2$^n$ & 17$\fs$8843(1) & 46$\farcs$508(1) & 0.515 & j \\
$-$11.4$^n$ & 17$\fs$6966(7) & 52$\farcs$311(6) &  0.195 & K & $\,\,\,-$8.0          & 17$\fs$9978(5) & 49$\farcs$678(4) & 0.131 & k \\
 $\,\,\,-$8.7$^n$ & 17$\fs$7184(6) & 49$\farcs$999(5) &  0.216 & L & $\,\,\,-$7.0     & 17$\fs$9084(5) & 50$\farcs$6132(5)& 0.090 & l \\
 $\,\,\,-$8.6     & 17$\fs$8852(1) & 46$\farcs$484(1) &  2.630 & M & $\,\,\,-$2.8$^n$ & 17$\fs$9985(4) & 49$\farcs$667(3) & 0.135 & m \\
 $\,\,\,-$7.9     & 17$\fs$9947(1) & 49$\farcs$682(1) &  5.192 & N & $\,\,\,-$0.6$^n$ & 17$\fs$9111(11)& 49$\farcs$361(7) & 0.423 & n \\
$\,\,\,-$7.9$^n$  & 17$\fs$8273(4) & 53$\farcs$190(3) & 0.471  & O & $\,\,\,-$0.6$^n$ & 17$\fs$9498(1) & 49$\farcs$843(1) & 15.370 & o \\
 $\,\,\,-$7.8$^n$ & 18$\fs$3843(12)& 47$\farcs$314(10)&  0.098 & P & $\,\,\,-$0.6$^n$ & 17$\fs$7829(3) & 53$\farcs$363(3) & 1.097 & p \\
 $\,\,\,-$7.7$^n$ & 17$\fs$9582(11)& 49$\farcs$175(6) &  0.209 & Q & &&&& \\
 $\,\,\,-$6.7$^n$ & 17$\fs$9478(2) & 49$\farcs$868(1) &  0.583 & R & &&&& \\
 $\,\,\,-$6.2     & 18$\fs$1163(1) & 49$\farcs$386(1) &  4.590 & S & &&&& \\
 $\,\,\,-$6.2$^n$ & 17$\fs$9488(4) & 52$\farcs$906(3) &  0.374 & T & &&&& \\
 $\,\,\,-$6.1$^n$ & 17$\fs$8903(7) & 46$\farcs$497(5) &  0.156 & U & &&&& \\
 $\,\,\,-$6.1$^n$ & 18$\fs$0778(9) & 48$\farcs$881(6) &  0.167 & V & &&&& \\
 $\,\,\,-$4.9     & 18$\fs$0497(2) & 47$\farcs$552(1) &  0.481 & W & &&&& \\
 $\,\,\,-$2.4$^n$ & 17$\fs$8407(2) & 49$\farcs$898(2) &  0.293 & Y & &&&& \\
 $\,\,\,-$0.7     & 17$\fs$9520(7) & 49$\farcs$848(5) &  0.109 & Z & &&&& \\
\hline
\multicolumn{5}{|l|}{1667-MHz LHC} & \multicolumn{5}{|l|}{1667-MHz RHC} \\
$-$15.7$^n$ & 18$\fs$3852(4) & 42$\farcs$737(3) & 0.936 & A &$-$14.6       & 18$\fs$1123(1) & 46$\farcs$306(1) &  3.962 & a \\
$-$15.7     & 18$\fs$1125(1) & 46$\farcs$305(1) & 4.831 & B & $\,\,\,-$5.4 & 17$\fs$8358(2) & 49$\farcs$912(2) &  0.611 & b \\
 $\,\,\,-$3.1$^n$ & 18$\fs$1096(4) & 46$\farcs$334(4) & 0.580 & C &      & & & &                               \\
 $\,\,\,-$3.1     & 17$\fs$8372(1) & 49$\farcs$899(1) & 2.907 & D &      & & & &                              \\
\hline
\multicolumn{4}{|l|}{1720-MHz LHC} & \multicolumn{4}{|l|}{1720-MHz RHC} \\
$-$13.3$^n$ & 17$\fs$6124(1) & 44$\farcs$533(1) & 1.2354 & A & $-$15.4$^n$ & 17$\fs$6147(1) & 44$\farcs$517(1) & 1.122 & a \\
\hline
\end{tabular}
\end{table*}

Spatially complex OH maser emission was found at 1665 and 1667\,MHz
in the velocity ranges from $-$26 to 0.1\,km\,s$^{-1}$ 
and from $-$16 to $-$3\,km\,s$^{-1}$, respectively.
The spectra of the LHC and RHC polarizations and the linear
polarization ($ P=(Q^2+U^2)^{0.5}$) are presented in Fig. 1.
No significant differences were noted in the spectra at the 
two epochs spanned by 10\,days. Our circularly polarized 
spectra differ considerably from those reported in \citet{cohen84}.  
However, Cep\,A is noted for OH flare activity and strong variability 
\citep{cohen85}.

No 1612-MHz emission was detected above a level of 15\,mJy over 
the whole region searched at both epochs. This is consistent with 
single dish observations by \cite*{cohen90} who found  weak and broad
1612-MHz emission which was resolved-out by MERLIN.

The 1720-MHz emission appeared on 1999 May 20 as completely circularly
polarized features with a peak intensity of about 1\,Jy (Fig.\/ 1) 
which decreased by a factor of two after four weeks \citep{niezurawska04}.
We did not detect any linear polarization at that transition 
within a sensitivity limit of 15\,mJy.

\subsection{Distribution of the maser components}
\begin{figure}
\resizebox{\hsize}{!}{\includegraphics{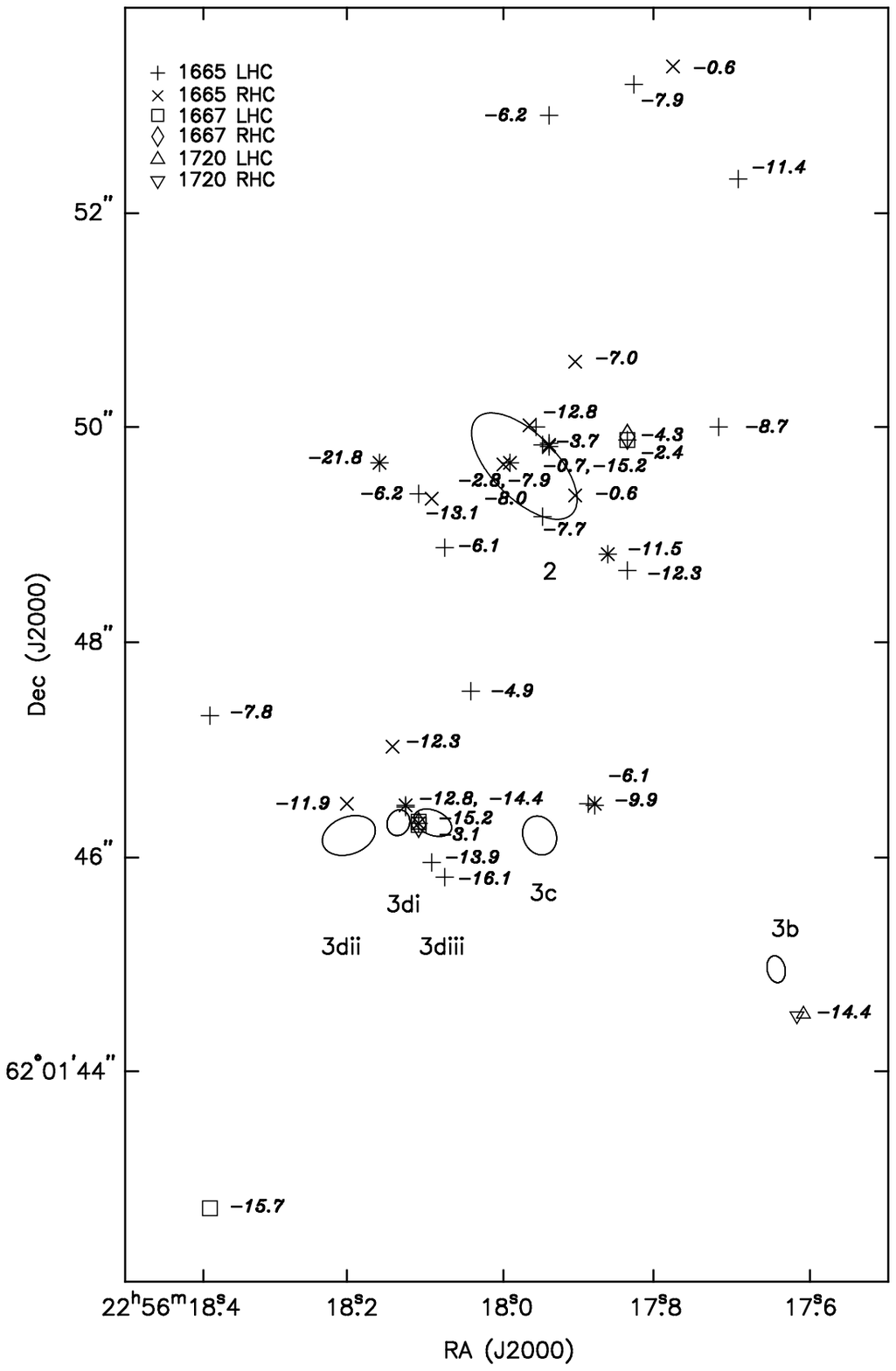}}
\caption{The OH maser distribution in the Cep\,A region measured with MERLIN. 
Symbols trace the 1665-, 1667- and 1720-MHz LHC and RHC features as indicated
in the top-left corner.
Each feature is labelled by its central LSR velocity (italic numbers). 
In a case of Zeeman pairs the demagnetized velocity is given.
Ellipses trace the continuum sources \citep{torrelles98}. 
Upright numbers label the continuum sources according 
to \citet{hughes84} and \citet{torrelles98}.} 
\label{fig2} 
\end{figure}

All OH maser features above 5$\sigma$ level are listed in Table 2 and
plotted in Fig. 2. In total, we detected 49 OH masers at 1.6\,GHz lines, 31
features are new (marked with superscript $n$ in Table 2) and 18 were seen
previously (\citealt{cohen84}; \citealt{migenes92}; \citealt*{argon00}).
 The 6 arcsec by 10 arcsec region contains
25 LHC and 16 RHC features in the 1665\,MHz line, four LHC and two RHC
features in the 1667\,MHz line and single LHC and RHC features in the
1720\,MHz line.

For the assumed distance of 725\,pc (\citealt*{blaauw59}) the
projected area containing OH maser emission is 4350$\times$7250\,au$^2$
i.e.\/ about twice as great as that reported previously.
The masers are mostly projected against H{\small II} regions labelled
as 2, 3b, 3c (the nomenclature of \citet{hughes84}) and 3di, 3dii,
3diii regions resolved by \citet{torrelles98}. Note that we use the
convention presented by \citet{torrelles98} in their Table 1 but not
in their Fig. 1.

24 OH maser features at 1665- and 1667-MHz are centred around the
H{\small II} region 2 within the velocity range from
$-$25.2\,km\,s$^{-1}$ to $-$0.6\,km\,s$^{-1}$. 14 OH features from
both transitions with velocities from $-$16.1\,km\,s$^{-1}$ to
$-$3.1\,km\,s$^{-1}$ were projected towards the continuum sources 3di,
3dii, 3diii. Three OH 1665-MHz features with velocities $-$11.2,
$-$8.6, $-$6.1\,km\,s$^{-1}$ were close the east edge of the H{\small
II} region 3c. Six features, which all were detected for the first
time, are not obviously associated with any known continuum
source. Four 1665-MHz masers with LSR velocities of $-$11.4, $-$7.9, $-$6.2
and $-$0.6\,km\,s$^{-1}$ are located 3.8 arcsec northward of
the continuum source 2.  One 1665-MHz feature with an LSR velocity
of $-$7.8\,km\,s$^{-1}$ lies about 2 arcsec NW of the 3dii centre,
while one 1667-MHz feature, at $-$15.7\,km\,s$^{-1}$, is placed
3.8 arcsec southward of the source 3dii (Fig. 2).  Two 1720-MHz
features are projected towards the southern edge of source 3b and
coincide with the OH 4765-MHz emission which we reported in
\citet{niezurawska04}.

\begin{table*}
\caption{1.6-GHz Zeeman pairs found in Cep A. 
The peak flux densities (S$_{\rm p}$), the separation in
velocity ($\Delta$V) between two features of opposite circular polarization,
the demagnetized velocities (V$_{\rm d}$) of those features 
and calculated magnetic field strengths (B) are given. The negative sign 
corresponds to the field towards the observer. The last column
identifies features according to the nomenclature from Table 1.}
\begin{tabular}{llcccccccc}
\hline
     &Polzn. & RA & Dec &  V$_{\rm LSR}$ & S$_{\rm p}$& $\Delta$V & V$_{\rm d}$ & B & \\
     && (22$^{\rm h}$56$^{\rm m}$) & (62\degr01\arcmin) &(km\,s$^{-1}$) 
& (Jy\,b$^{-1}$)& (km\,s$^{-1}$) & (km\,s$^{-1}$) & (mG)  & \\
\hline
1665-MHz  &&&&&&&& \\
           &LHC & 17\fs9651  & 50\farcs006 &$-$13.9 & 1.426 &$-$2.2 & $-$12.8 &  3.7 & H \\
{\bf Z$_1$}&RHC & 17\fs9699  & 50\farcs012 &$-$11.7 & 0.130 &       & & & g     \\
&&&&&&&& \\
           &LHC & 17\fs9478  & 49\farcs868 &$-$6.7  & 0.583 &$-$6.1 & $-$3.7  & 10.3 & R \\
{\bf Z$_2$}&RHC & 17\fs9498  & 49\farcs843 &$-$0.6  &15.370 &       & & & o \\ 
&&&&&&&& \\
           &LHC & 17\fs9472  & 49\farcs829 &$-$16.1 & 0.779 &$-$1.9 & $-$15.2 & 3.2 & C  \\
{\bf Z$_3$}&RHC & 17\fs9474  & 49\farcs839 &$-$14.2 & 0.360 &       & &  & c   \\
&&&&&&&& \\
           &LHC & 18\fs1642  & 49\farcs668 &$-$25.5 & 0.285 &$-$7.5 & $-$21.8 & 12.7 & A  \\
{\bf Z$_4$}&RHC & 18\fs1646  & 49\farcs673 &$-$18.0 & 0.174 &       & &    & a  \\
&&&&&&&& \\
           &LHC & 17\fs8852  & 46\farcs484 &$-$8.6  & 2.630 & 2.6   & $-$9.9 & $-$4.4 & M   \\ 
{\bf Z$_5$}&RHC & 17\fs8843  & 46\farcs508 &$-$11.2 & 0.515 &       & & & j   \\
&&&&&&&& \\
           &LHC & 18\fs1324  & 46\farcs494 &$-$13.9 & 16.250 &$-$2.3 & $-$12.8 &  3.9 & F \\
{\bf Z$_6$}&RHC & 18\fs1352  & 46\farcs487 &$-$11.6 & 2.070  &       & &  & h  \\ 
&&&&&&&& \\
           &LHC & 18\fs1148  & 46\farcs320 &$-$16.1 & 8.880 &$-$1.9 & $-$15.2 & 3.2 & D  \\ 
{\bf Z$_7$}&RHC & 18\fs1143  & 46\farcs316 &$-$14.2 & 5.240 &       & & & b   \\
&&&&&&&& \\
\hline
1667-MHz  &&&&&&&& \\
           &LHC & 17\fs8372 & 49\farcs899 &$-$3.1 & 2.907  &  2.3  & $-$4.3 & $-$6.5 & D \\ 
{\bf Z$_8$}&RHC & 17\fs8358 & 49\farcs912 &$-$5.4 & 0.611  &       &  & & b    \\
&&&&&&&& \\
           &LHC & 18\fs1125 & 46\farcs305 &$-$15.7  & 4.831 &$-$1.1 & $-$15.2 & 3.1 & B \\ 
{\bf Z$_7$}&RHC & 18\fs1123 & 46\farcs306 &$-$14.6  & 3.962 &       & & & a    \\
&&&&&&&& \\
\hline 
1720-MHz &&&&&&&& \\
     	   &LHC & 17\fs6124 & 44\farcs533 &$-$13.3 & 1.050 & 2.1   & $-$14.4 & $-$17.3 & A \\
{\bf Z$_9$}&RHC & 17\fs6147 & 44\farcs517 &$-$15.4 & 1.122 &       &  && a   \\
\hline
\end{tabular}                                
\end{table*} 

\subsection{Polarization properties and magnetic field}
We found ten Zeeman pairs with features coinciding spatially 
within 36\,mas. Table 3 summarizes their properties.
Seven pairs were found at 1665\,MHz, two at 1667\,MHz and one at 1720\,MHz. 
Five of them, named Z$_1$, Z$_2$, Z$_3$, Z$_4$ and Z$_8$ 
are located around the H{\small II} region 2. Three pairs, Z$_6$, Z$_7$ (1665\,MHz) and
Z$_7$ (1667\,MHz), are projected at the sources 3di and 3diii. Pair Z$_5$
lies close to the region 3c, whereas the only one Zeeman pair at 1720\,MHz line 
appears in the vicinity of the H{\small II} region 3b.
We calculated the magnetic field strength
(B) from the velocity separation of the Zeeman components
($\Delta$V) using coefficients taken from Table 2 in \citet{davies74}.
However, as we stated in \citet{niezurawska04},
in the case of 1720-MHz transition we assumed that the
emission came from well$-$separated $\sigma^{+1}$ and $\sigma^{-1}$
components and used the line splitting of 118\,km\,s$^{-1}$ G$^{-1}$.

Two Zeeman pairs, one at 1665\,MHz and one at 1667\,MHz, 
labelled as Z$_7$, coincide spatially and 
in velocity. In the past they were reported by \citet*{wouterloot80} 
as a Zeeman quartet. Channel maps of that quartet in linearly polarized 
emission are shown in Fig.\,3 and the polarization profiles in Fig.\,4.  
The emission at $-$15.8\,km\,s$^{-1}$ 
appears at both transitions with similar linear polarized flux densities,
$P\approx$145\,mJy\,b$^{-1}$, while the polarization angles differ
significantly by 23\degr. We think that blending of features unresolved by
MERLIN is the most likely explanation for that difference (see Fig.\,3 and
compare e.g.\,panels V$_{\rm LSR}$=$-$14.7\,km\,s$^{-1}$ at both transitions). 
Moreover, the single dish spectra of the quartet showed asymmetries in the line
profiles, i.e.\,more than one Gaussian component was needed for an accurate fit
\citep{cohen90}. 

\begin{figure*}
        \centering
\resizebox{\hsize}{!}{\includegraphics{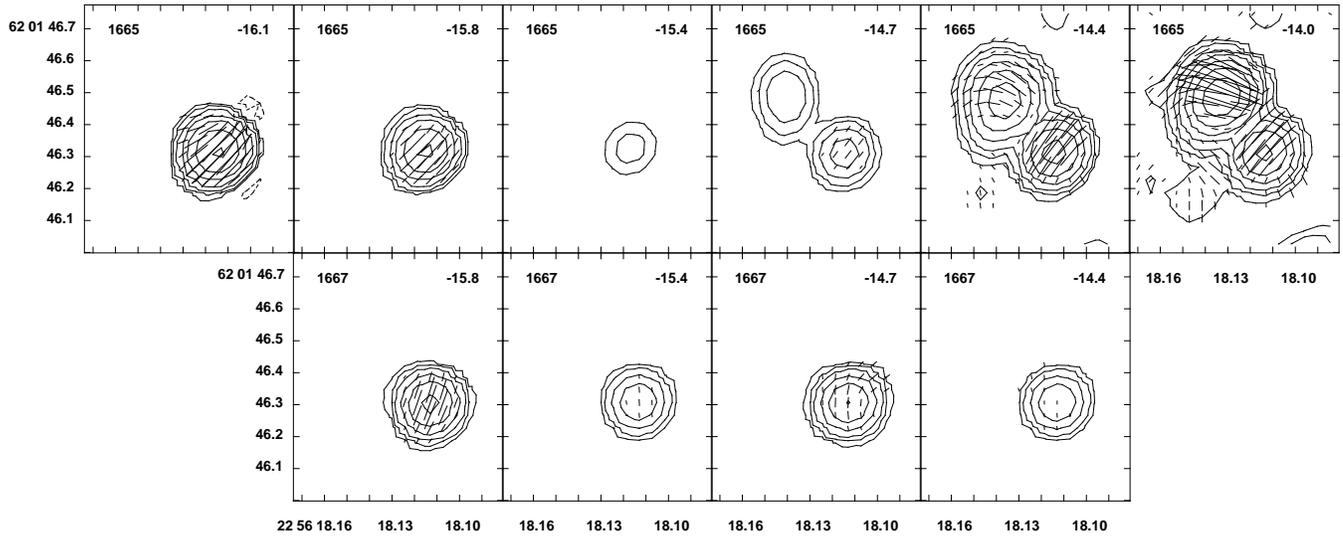}}
\caption{Channel maps of linear polarization of the Zeeman quartet at
1665 and 1667\,MHz. Contour levels are $-$1, 1, 2, 4,
...$\times$60\,mJy\,b$^{-1}$. LSR velocities are indicated
at the upper$-$left corners. Bars show the orientation of the electric
vectors and have lengths proportional to the linearly polarized intensities
where 0.1 arcsec=125\,mJy\,b$^{-1}$.}
\label{fig3} 
\end{figure*}

\begin{figure}
\resizebox{\hsize}{!}{\includegraphics{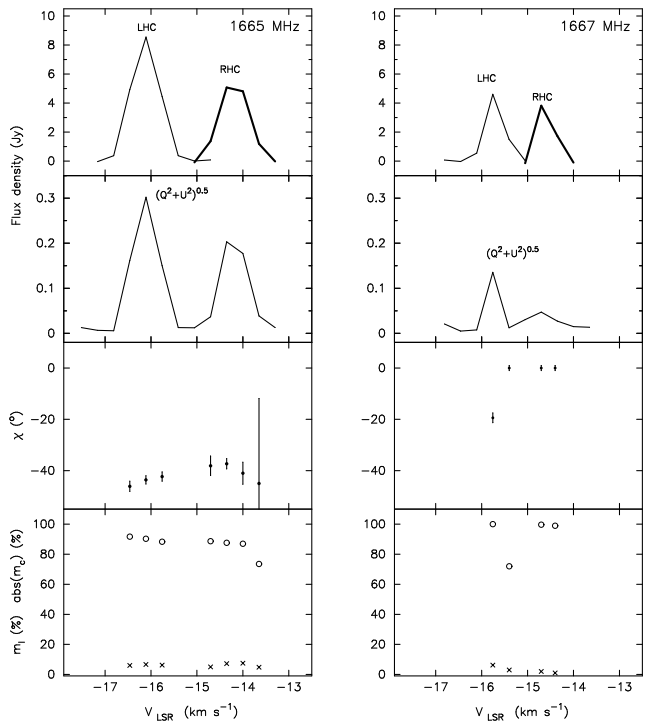}}
\caption{Polarization profiles of the OH Zeeman quartet in Cep\,A at 1665 and
1667\,MHz. In the lowest panel the circle and cross symbols correspond to
$m_{\rm c}$ and $m_{\rm l}$, respectively.}
\label{fig4}
\end{figure}

Table 4 gives the polarization characterictics of all maser features
detected in 1999; the LSR
velocity, the J2000 coordinates together with their errors, the $I, Q,
U, V$ and $P$ flux densities, the polarization position angle
$\chi=0.5\times\arctan(U/Q)$ and its error $\sigma_\chi$, 
the percentage of linear
polarization, $m_{\rm l}=(Q^2+U^2)^{0.5}/I$, the percentage of
circular polarization, $m_{\rm c}=V/I$ and the percentage of total
polarization, $m_{\rm t}=(m_{\rm c}^2 + m_{\rm l}^2)^{0.5}$.  
The random errors $\sigma_\chi$ due to thermal noise
were calculated according to the formula $\sigma_\chi=
0.5\times \sigma_P/P \times 180\degr / \pi$ \citep{wardle74}.
The errors arising from uncertainty in the measurement of the
polarization angle of 3C286 were 4\degr and 3\degr\, at 1665 and
1667\,MHz respectively.  These systematic errors do not affect
comparisons within the same data set, so they are not included in
Table 4.

We found 14 linearly polarized OH maser features at 1665 MHz and two at 
1667 MHz  above the 5$\sigma$ level ($P\ge0.010$\,mJy).
 Even when linear polarization is detected 
in Cep\,A the percentage polarization is generally low ($\le$16\,per\,cent); 
we detected 
only two 1665-MHz components with $m_{\rm l}$ of 22 and 31\,per\,cent. 
In contrast, the circular polarization percentages are typically 
higher with a median value of 77\,per\,cent and with two cases fully
polarized, in 1665\,MHz line. Similary, the 1667-MHz line shows
$\le$6\,per\,cent linear polarization but almost full circular polarization
with a median value $m_{\rm c}$=90\,per\,cent. 
The distribution of linearly polarized components together with their
polarization characteristic are presented in Fig.~5.
We did not find any linear polarization in the 1720-MHz transition above the
5$\sigma$ limit. 

\begin{table*}
\caption{The linearly polarized OH 1665-MHz and 1667-MHz maser components  
in Cep\,A. The errors are given in brackets and are $\times$0.0001 sec for RA,
$\times$0.001 arcsec for Dec and in degrees for $\chi$. The last column
identifies features according to the nomenclature from Table 1.}
\begin{tabular}{p{8mm}p{14mm}p{12mm}p{9.5mm}p{9.5mm}p{9.5mm}p{9.5mm}p{12mm}p{8mm}p{5.5mm}p{5.5mm}p{5.5mm}p{5.5mm}}
\hline
$V_{\rm LSR}$ & RA & Dec & \hspace*{5mm}$I$ & \hspace*{5mm}$Q$ & \hspace*{5mm}$U$ & \hspace*{5mm}$V$ & \hspace*{5mm}$P$ & \hspace*{1mm}$\chi$ & m$_{\rm l}$ & 
m$_{\rm c}$ & m$_{\rm t}$\\
(km\,s$^{-1}$) &(22$^{\rm h}$56$^{\rm m}$) & (62\degr01\arcmin) &
(Jy\,b$^{-1}$)& (Jy\,b$^{-1}$)& (Jy\,b$^{-1}$)&
(Jy\,b$^{-1}$)& (Jy\,b$^{-1}$)& ($^{o}$) & (\%) & (\%) & (\%) & \\
\hline
\end{tabular}
\begin{tabular}{rllcrrrrrrrrr}
\multicolumn{13}{l}{1665-MHz} \\
$-$25.3 & 18\fs1647(4) & 49\farcs667(3) & 0.141 & $-$0.031 & $-$0.006 & $-$0.132 & 0.031 &$-$85(3)  & 22 & $-$94 & 97 & A \\ 
$-$18.0 & 18\fs1646(3) & 49\farcs673(3) & 0.083 & 0.002    &  0.002   & 0.066    & 0.003    & \multicolumn{1}{c}{$-$}   & 0  & 80 & 80 & a \\ 
$-$16.1 & 17\fs9471(6) & 49\farcs827(4) & 0.393 & $<$0.002 & $<$0.002 & $-$0.337 & $<$0.002 & \multicolumn{1}{c}{$-$}   & 0  & $-$86 & 86 & C \\
$-$16.1 & 18\fs0773(17)& 45\farcs804(12)& 0.172 & $<$0.002 & $-$0.018 & $-$0.091 & 0.018 &$-$45(3) & 11 & $-$53 & 54 & B \\
$-$16.1 & 18\fs1145(1) & 46\farcs318(2) & 4.483 &  0.015   & $-$0.293 & $-$4.049 & 0.294 &$-$44(1) & 7  & $-$90 & 90 & D \\
$-$14.4 & 18\fs1138(1) & 46\farcs313(2) & 2.661 &  0.051   & $-$0.184 &  2.331   & 0.191 & $-$37(2)  & 7  & 88    & 88 & b \\
$-$14.0 & 17\fs9654(4) & 50\farcs313(4) & 0.678 & $<$0.002 & $<$0.002 & $-$0.491 & $<$0.002 & \multicolumn{1}{c}{$-$}   & 0  & $-$72 & 72 & H \\
$-$14.0 & 18\fs0949(8) & 45\farcs985(9) & 0.334 & $<$0.002 & $<$0.002 & $-$0.106 & $<$0.002 & \multicolumn{1}{c}{$-$}   & 0  & $-$31 & 31 & G \\
$-$14.0 & 18\fs1420(13)& 46\farcs426(34)& 0.708 & $-$0.046 & $<$0.002 & $-$2.420 & 0.046 & 90(2)   & 7  & $-$100& 100 &  E \\
$-$13.6 & 18\fs1309(1) & 46\farcs494(2) & 6.923 & $-$0.274 & 0.179    & $-$6.176 & 0.328 & 73(2)     & 5  & $-$89 & 90& F \\
$-$13.0 & 18\fs0979(2) & 49\farcs346(2) & 0.182 & 0.004    & $-$0.006 & 0.150    & 0.007 & \multicolumn{1}{c}{$-$}      & 0  &  83   & 83 & d \\
$-$12.6 & 18\fs1459(7) & 47\farcs029(6) & 0.101 & $-$0.006 & $-$0.015 & 0.056    & 0.016 & $-$55(4)& 16 & 55    & 57 & e \\
$-$12.2 & 17\fs8379(2) & 48\farcs674(2) & 0.314 & $<$0.002 & $-$0.003 &$-$0.233  &  0.003   & \multicolumn{1}{c}{$-$}   & 0  & $-$74 & 74 & I \\  
$-$11.9 & 17\fs9705(7) & 50\farcs032(9) & 0.058 & $<$0.002 & $<$0.002 & 0.040    & $<$0.002 & \multicolumn{1}{c}{$-$}   & 0  & 68    & 68 & g\\
$-$11.9 & 18\fs1379(2) & 46\farcs499(2) & 1.019 & $<$0.002 & $-$0.021 & 0.763    & 0.021 & $-$45(5)  & 2  & 75    & 75 & h \\
$-$11.5 & 17\fs8657(2) & 48\farcs810(2) & 1.225 & $<$0.002 & $<$0.002 & $-$0.775 &$<$0.002& \multicolumn{1}{c}{$-$}     & 0  & $-$63 & 63 & J \\
$-$11.2 & 17\fs8845(2) & 46\farcs507(2) & 0.240 & 0.009    &$<$0.002  &   0.207  & 0.009 & \multicolumn{1}{c}{$-$}      & 0  & 86    & 86 & j \\
 $-$8.7 & 17\fs8848(1) & 46\farcs487(2) & 1.354 & $-$0.013 &$<$0.002  &$-$1.094  & 0.013 & 90(5)   & 1  & $-$81 & 81 & M \\
 $-$8.0 & 17\fs9953(1) & 49\farcs681(2) & 2.681 & 0.086    & $-$0.100 &$-$2.208  & 0.132 & $-$25(1)  & 5  & $-$82 & 82 & N \\
 $-$8.0 & 17\fs9593(20)& 49\farcs172(14)& 0.079 & $<$0.002 & $-$0.009 &$-$0.059  & 0.009 & \multicolumn{1}{c}{$-$}      & 0  & $-$75 & 75 & Q \\ 
 $-$8.0 & 17\fs8275(5) & 53\farcs189(6) & 0.229 & $<$0.002 & $<$0.002 & $-$0.195 & $<$0.002 & \multicolumn{1}{c}{$-$} & 0 & $-$85 & 85 & O \\
 $-$7.0 & 17\fs9085(9) & 50\farcs648(8) & 0.131 & $-$0.041 & $-$0.004 &$<$0.002  & 0.041 &$-$87(2)& 31 & 0     & 31 & l \\
 $-$6.6 & 17\fs9478(2) & 49\farcs867(2) & 0.300 & $-$0.005 & $<$0.002 & $-$0.285 & 0.005 & \multicolumn{1}{c}{$-$} & 0 & $-$95 & 95 & R \\
 $-$6.3 & 18\fs1167(1) & 49\farcs383(2) & 2.326 & $-$0.035 &  0.037   & $-$1.884 & 0.050 &67(2) & 2 & $-$81 & 81& S \\  
 $-$6.3 & 18\fs0788(10)& 48\farcs872(10)& 0.091 & $<$0.002 & $<$0.002 & $-$0.072 & $<$0.002  & \multicolumn{1}{c}{$-$} & 0 & $-$79 & 79 & V \\
 $-$6.3 & 17\fs9492(6) & 52\farcs905(7) & 0.190 & $<$0.002 & $<$0.002 & $-$0.142 & $<$0.002  & \multicolumn{1}{c}{$-$} & 0 & $-$75 & 75 & T \\
 $-$5.9 & 17\fs8907(8) & 46\farcs496(9) & 0.088 & $<$0.002 & $<$0.002 & $-$0.058 & $<$0.002 &  \multicolumn{1}{c}{$-$} & 0 & $-$66 & 66 & U \\
 $-$4.9 & 18\fs0496(2) & 47\farcs550(2) & 0.233 &$<$0.002  &$<$0.002  & $-$0.211 & $<$0.002 &  \multicolumn{1}{c}{$-$} & 0 & $-$90 & 90 & W \\
 $-$2.4 & 17\fs8410(3) & 49\farcs899(3) & 0.146 &  0.003   & $<$0.002 & $-$0.127 &    0.003 &  \multicolumn{1}{c}{$-$} & 0 & $-$87 & 87 & Y \\ 
 $-$1.0 & 17\fs9122(24)& 49\farcs354(15)& 0.089 &$<$0.002  & $-$0.010 &  0.102   &  0.010 &$-$45(6)&11&  100 & 100 & n \\
 $-$0.6 & 17\fs9499(5) & 49\farcs842(5) & 7.761 & $-$0.025 & $-$0.102 &  6.852   & 0.106 &$-$52(1)   & 1  & 88    & 88 & o \\
 $-$0.6 & 17\fs7829(4) & 53\farcs362(5) & 0.535 & $<$0.002 & $<$0.002 &  0.442  &$<$0.002 & \multicolumn{1}{c}{$-$} & 0 & 83 & 83 & p \\
\multicolumn{13}{l}{1667-MHz} \\
$-$15.8 & 18\fs1125(1) & 46\farcs303(2) & 2.324 & 0.111    & $-$0.090 & $-$2.422 & 0.143 &$-$19(1)  & 6 & $-$100 & 100 & B \\
$-$15.7 & 18\fs3852(4) & 42\farcs737(3) & 0.482 & $<$0.002 &$<$0.002  & $-$0.455 &$<$0.002 & \multicolumn{1}{c}{$-$}    & 0 & $-$94  & 94 & A \\
$-$14.7 & 18\fs1123(1) & 46\farcs306(2) & 1.937 & 0.039    &$<$0.002  &  1.932   & 0.039 & 0(2)      & 2 & 100    & 100 & a \\
$-$5.4  & 17\fs8358(2) & 49\farcs912(2) & 0.334 &$<$0.002  &$<$0.002  &  0.281   & $<$0.002 & \multicolumn{1}{c}{$-$}   & 0 & 84 & 84 & b \\ 
$-$3.1  & 18\fs1096(4) & 46\farcs334(4) & 0.397 &$-$0.008  & $<$0.002 &$-$0.261  & 0.008    & \multicolumn{1}{c}{$-$}   & 0 & $-$66  & 66 & C \\  
$-$3.1  & 17\fs8372(1) & 49\farcs899(1) & 1.481 &$<$0.002  &$<$0.002  &$-$1.412  & $<$0.002 & \multicolumn{1}{c}{$-$}   & 0 & $-$95 & 95 & D \\
\multicolumn{13}{l}{1720-MHz} \\
$-$15.4 & 17\fs6147(1) & 44\farcs517(1) & 0.531 &$<$0.002  &$<$0.002  & 0.525    & $<$0.002 & \multicolumn{1}{c}{$-$}   & 0 & 99 & 99 & a \\
$-$13.3 & 17\fs6124(1) & 44\farcs533(1) & 0.507 &$<$0.002  &$<$0.002  & 0.561    & $<$0.002 & \multicolumn{1}{c}{$-$}   & 0 & $-$100 & 100
& A \\
\hline
\end{tabular}                                
\end{table*} 

\section{Discussion}
\subsection{OH masers and radio continuum sources}
The present study reveals that the OH masers are spread over an area
of 6 arcsec by 10 arcsec. In addition to two main clusters of maser
components (\citealt{cohen84}) associated with the radio continuum
sources 2 and 3 (\citealt{hughes84}) we detected several components
which have not been reported before and which are not accompanied by
radio continuum background. 94\,per\,cent of the new components are
much brighter than the 0.1\,Jy sensitivity limit of previous
observations, demonstrating the high variability of OH masers.

OH 1665-MHz masers around region 2 (which has a complex structure)
encompass a wider range of velocities ($-$25.2\,km\,s$^{-1}$ to
$-$0.6\,km\,s$^{-1}$) than that reported by \citet{cohen84}. This
maser complex extends 3 arcsec (2175\,au) in an E--W direction and its
geometrical centre is at RA=22$^{\rm h}$56$^{\rm m}$17\fs9609,
Dec=62$\degr01'49\farcs578$ (J2000). That gives a distance of 147\,mas
(107\,au) from the geometrical centre of a disc traced by water masers
\citep{torrelles96}. Although the E--W elongations of the OH and the
water masers are similar, the position$-$velocity diagram for OH
masers (after demagnetization of the velocities to compensate for
Zeeman splitting) does not show the regularity expected from a disc
scenario. 
Our results imply that the existing disc has to be
strongly disturbed at the distances 1-2 arcsec from the centre of
region 2 by the expanding material.  In fact, there are molecular
outflows at the edge (about 1 arcsec from the centre) of region 2
\citep{migenes92}.

The OH 1665- and 1667-MHz masers projected against the southern continuum 
source 3d cover mainly a velocity range from $-$16.1\,km\,s$^{-1}$ to 
$-$11.6\,km\,s$^{-1}$. However, we found a highly red-shifted component
of velocity $-$3.1\,km\,s$^{-1}$. \citet{torrelles98} resolved 3d into a chain 
of four individual 22\,GHz continuum sources. They point to 3dii as harbouring 
a young  stellar object. We only detected OH maser emission from the west side 
of this  source.  If this was powered by an outflow we would expect to see 
a velocity gradient or a systematic spatial distribution of the masers but 
this was not apparent even after demagnetization of the component velocities.
\citet{torrelles98} reported  a similar situation for the water masers around 
that region, there were neither clear spatial nor velocity trend in
distribution of maser spots. 

\subsection{Magnetic field}
\begin{figure*}
\resizebox{\hsize}{!}{\includegraphics{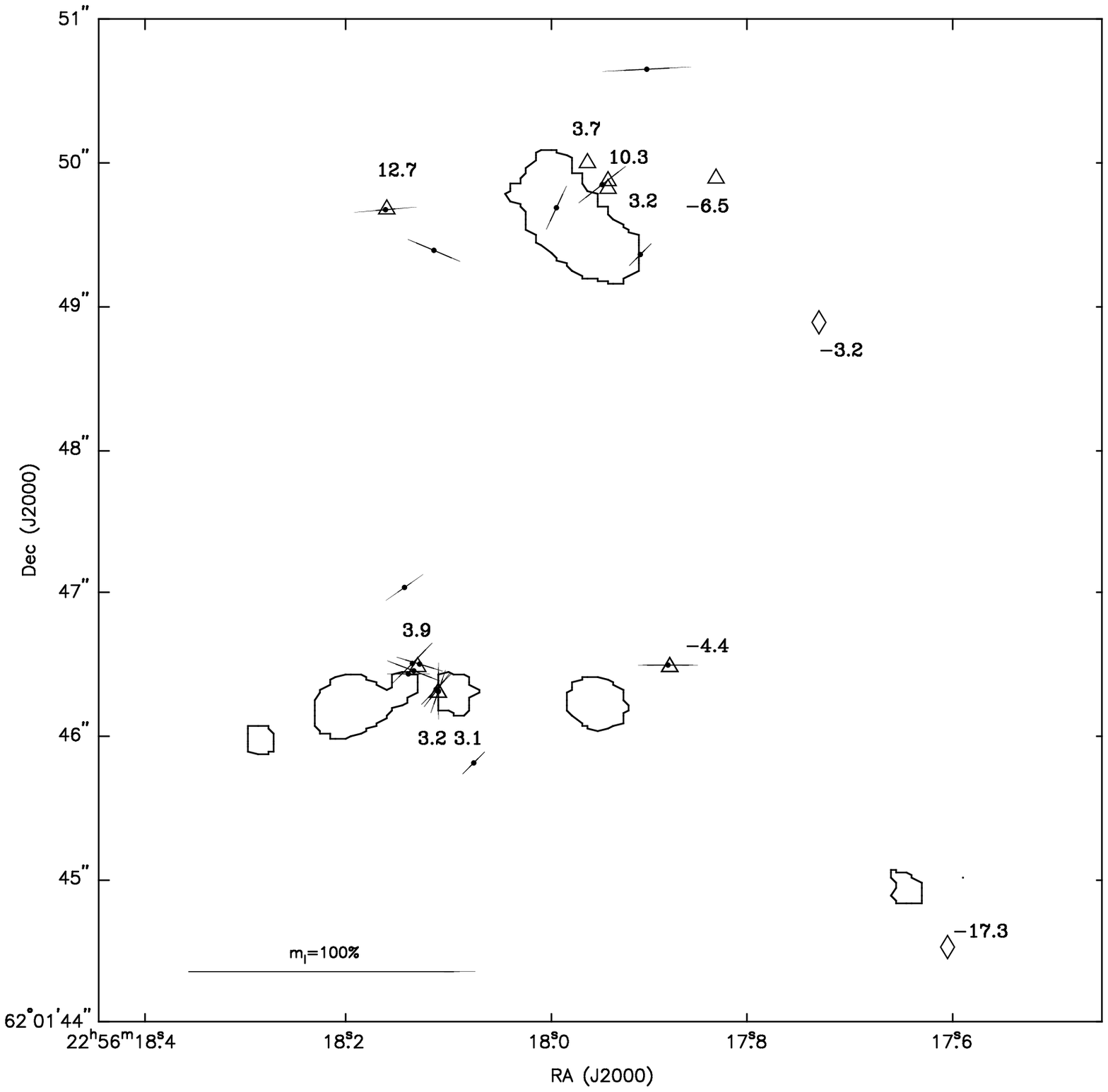}}
\caption{An overview of the magnetic field in Cep\,A detected using MERLIN. 
 Numbers indicate the magnetic field strength in mG. 
The directions of the planes containing the electric field vectors are marked 
by bars with lengths  proportional to the percentage linear polarization 
of each feature. Triangles trace the OH 1665- and
1667-MHz Zeeman pairs from this paper. Diamonds show the water maser Zeeman
pair from \citet{sarma02} and OH 1720-MHz Zeeman pair 
\citep{niezurawska04}.
The contours are at 0.5\,mJy\,b$^{-1}$ (3\,per\,cent of the peak intensity) 
of the 22-GHz continuum emission \citep{torrelles98}.} 
\label{fig5} 
\end{figure*}

The most important result of our polarimetric observations concerns
the overall regularity of the magnetic field in Cep\,A. Fig.~5 shows
the distribution of maser components with polarization information.
The magnetic field, derived from the Zeeman splitting of OH 1665- and
1667-MHz lines, is negative (pointed towards us) in the west and
positive (pointed away from us) in the east side of the mapped region.
\citet{sarma02} detected a Zeeman pair in the water maser line at
22\,GHz and derived an upper limit to the field strength of
$-$3.2\,mG.  That component lies just on the west side of Cep\,A
and is consistent with the magnetic field configuration inferred from OH
masers (Fig.~5). Furthermore, a Zeeman pair found in the 1720-MHz OH line
\citep{niezurawska04} in the SW region of the mapped area supports
this picture of an organised magnetic field on arcsecond scales. \\

The magnetic field reversal in Cep\,A is clearer than the reversals found  
in the molecular outflow sources studied with MERLIN by Hutawarakorn 
\& Cohen (1999, 2003, 2005) and \citet{hutawarakorn02}.  
However it is more difficult to relate the organised magnetic field 
structure seen in Cep\,A on the 1~arcsec scale to the diverse outflows 
seen on larger scales from $\sim$1 to 3~arcmin.  
Table~5 lists the main molecular structures reported towards Cep\,A.   
There are four preferred directions:  $-$45\degr, 0\degr, $+$45\degr and 
$+$90\degr.  Only H$_{2}$S shows an outflow at 0\degr\, in the expected 
direction orthogonal to the magnetic field reversal according to the model
by \citet{uchida85}. The multiplicity 
of outflow directions suggests that there may be several sources 
powering them.  The present OH maser 
observations suggest that Cep\,A contains at least two active centres.  
There is a clear need for sensitive molecular observations at the 
arcsec scale which could disentangle the outflows from the complex 
region at the core of Cep\,A.

\begin{table*}
\caption {Structures in Cep\,A regions.}
\begin{tabular}{lclcll}
\hline
Line & Scale & P.A. & Synthesised & Note & Reference  \\
     & (arcsec) & (\degr) & beam (\arcsec) &   &      \\
\hline
NH$_3$ & 42$\times$30 & $-$45 & 5.4$\times$5.3 & &\citet{torrelles93} \\
       & 96$\times$24 & $+$45 & & & \\
       & 48$\times$36 & $-$135& & & \\
$^{12}$CO& 300 & $+$90   & 9.9$\times$6.3 & HV outflow & \citet{narayanan96} \\
         & 180 &   $+$85 & & EHV outflow & \\
       CS&  91 &   $-$45 & & core & \\
HCO$^+$  &  60 &   $+$57 & 3.2$\times$2.6 & outflow & \citet{gomez99} \\         
SiO      & 120 &   $-$33 & 4.7$\times$3.7 & disc & \\
H$_2$S   &  85 & $+$135 & 14 & HV outflow &\citet{codella03} \\
         & 171 & $+$180  & & HV outflow & \\
         &$\sim$60&  $+$45 & & LH outflow & \\
SO$_2$   & 114$\times$29& $+$45 & 24 & \\
\hline
\end{tabular}
\end{table*}

The electric vectors of linearly polarized features show a systematic
trend.  The polarization position angles $\chi$ of components observed
in the direction of the outer parts of the H{\small II} regions are
significantly rotated relative to those observed towards the centres
(Fig.~5).  The flux-averaged $\chi$ of the six electric vectors from
the outer parts is $-$87\degr$\pm$3\degr, which implies that the
component of the magnetic field vector in the plane of the sky has a
direction of 3\degr$\pm$3\degr. That is closely aligned with the
interstellar magnetic field orientation ($-$10\degr) derived from
the infrared polarimetry towards the regions 2 and 3
\citep*{jones04}.  
The flux averaged $\chi$ from the centre of northern region
2 is $-$37\degr$\pm$1\degr, which gives 53\degr$\pm$1\degr for the
magnetic field orientation in that area. That is parallel to the radio
jet seen at 22\,GHz (P.A.=44\degr) \citep{torrelles96}.  The flux averaged $\chi$ of
1665-MHz features in southern regions 3di and 3diii (labels as in Fig.~2)
are 78\degr$\pm$2\degr and $-$41\degr$\pm$2\degr, respectively. Those
imply average magnetic field directions of $-$12\degr$\pm$2\degr and
49$\pm$2\degr, respectively.

\subsubsection{Physical conditions}
The magnetic field and the gas density relationship in Cep~A region was
established to be B$\sim$n$^{0.4}$ \citep{garay96}. 
At the edge of NH$_3$ molecular clouds in Cep~A,
\citet{garay96} found a magnetic field strength of 0.3\,mG and a
number density of 2$\times$10$^4$\,cm$^{-3}$. In this paper, we find
that the magnetic field strength deduced from OH Zeeman splitting has
a magnitude of 3.2 to 17.3\,mG.  This leads to derived gas densities
in the range from 8$\times$10$^6$ to 5$\times$10$^8$\,cm$^{-3}$.  Such
numbers are typical for OH masers in star-forming regions, according
to the model of \citet*{gray91}.  Additionally, this model predicts that in the
case of accelerative fields with  velocity shifts of
2--3\,km\,s$^{-1}$ the 1665-MHz line dominates at all number
densities. In fact, the masing region in Cep~A expands with a velocity of
2.5\,km\,s$^{-1}$ \citep{cohen90} and the 1665-MHz
emission does appear at least twice as strong as the 1667-MHz
line (Fig.~1).

\citet{hutawarakorn03} suggested that the evolutionary path of
star-forming regions is seen in their polarization
characteristics. They detected a systematic increase in the degree of
polarization from the oldest source to the youngest one in NGC\,7538.
In general, Cep\,A showed highly polarized masers with weak linear
polarization (typically a few \,per\,cent) but strong circular polarization
(typically $>$50\,per\,cent). The median $m_{\rm t}$ for 1665- and
1667-MHz features is 80\,per\,cent. That confirms the young evolutionary stage
of the Cep\,A region (similarly to the object IRS 11 in NGC\,7538). There are no
significant differences between the polarization properties of regions 2 and
3 what implies a similar age for all H{\small II} regions.

\subsubsection{Decay of magnetic field}
\citet{cohen90} monitored the Zeeman quartet at 1665- and 1667-MHz
over a 10-year time span. They present evidence for a
0.4\,per\,cent\,yr$^{-1}$ decay of the magnetic field strength in the
OH maser region due to expansion of molecular gas surrounding a young
star.  In Fig.~6 we present measurements of the velocity separations
of the components of the Zeeman quartet over a longer period of 25
years. We include previously reported data from 1980$-$90
(\citealt{wouterloot80}; \citealt{cohen90}; \citealt{fish03}) and
unpublished measurements taken with the Lovell antenna in the period
1990$-$95 (Cohen et al., in prep.).  Our interferometric data gave
separations between peaks of 1.9\,km\,s$^{-1}$ at 1665\,MHz and
1.1\,km\,s$^{-1}$ at 1667\,MHz in 1999.  We also detected the Zeeman
quartet with the Nan\c{c}ay antenna in 2002 October and 2004 September
(Szymczak et al., in prep.). The velocity separations were then 1.952,
1.93\,km\,s$^{-1}$ (at 1665\,MHz) and 1.15, 1.11\,km\,s$^{-1}$ 
(at 1667\,MHz) in both epochs, respectively. 
Taking the whole 25-year period we used least-squares fits to estimate
the rates of change in line splitting, obtaining
$-0.0053\pm0.0006$\,km\,s$^{-1}$\,yr$^{-1}$ for the 1665-MHz line and
$-0.0013\pm0.0007$\,km\,s$^{-1}$\,yr$^{-1}$ for the 1667-MHz line.
The rates of change of field strength implied are 
$-0.009\pm0.001$\,mG\,yr$^{-1}$ from the 1665-MHz line 
and $-0.004\pm0.002$\,mG\,yr$^{-1}$ from the 1667-MHz line. 

The two estimates for the magnetic field decay differ significantly of the
2.3$\sigma$ level (3 per\,cent). This is also apparent in the lowest panel
of Fig.~6. The ratio of line splitting should be 1.67 from theory whereas the
observed ratios are 1.70$-$1.82.
It is most likely those deviations are caused by blending (Sect.\,3.3). 
In particular, 1665-MHz features C and c and 1667-MHz feature A all have
similar velocities to the Zeeman quartet. It is imposible to correct data
for this effect. We take the weighted mean of the 1665- and 1667-MHz results
as our best estimate of the magnetic field decay. This gives a  
weighted mean of the  magnetic field decay rate of
$-0.0080\pm0.0009$\,mG\,yr$^{-1}$ which is $-0.24\pm0.03$\,per\,cent\,yr$^{-1}$.

\begin{figure}
\includegraphics[width=70mm]{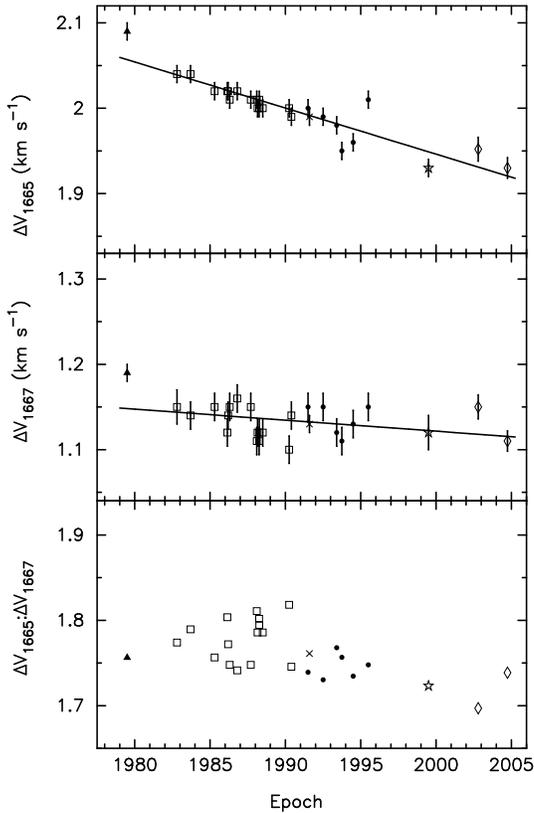}
\caption{The systematic decrease of the magnetic field in Cep\,A shown
by Zeeman quartet investigations. The open star represents data from this
paper taken with MERLIN and diamonds represent data taken with the 
Nan\c{c}ay telescope in 2002 and 2004 (Szymczak et al., in prep.). The 1979 data are from \citet{wouterloot80} (triangle), 
the 1982$-$1990 data are from \citet{cohen90} (squares), 
the 1990$-$1995 data were taken with the Lovell antenna (Cohen, in prep.)
(filled circles). A cross indicates the VLA data from \citet{fish03}. 
Bars indicate errors in measurements. For the details of fitting see
text.}
\label{fig6}
\end{figure}

\section{Conclusions}
We detected many new 1.6\,GHz OH masers in the Cep\,A region, 
and confirm the strong variability of the OH maser emission.  
OH maser features around H{\small II} region 2 have an elongated
E-W distribution, similar to that of the water masers. However, 
the OH maser kinematics do not show the regular position-velocity 
pattern seen in the water masers.  

The most important result from our full polarimetry observations 
is the morphology of magnetic field in close surroundings of the young
stars. The Zeeman pairs reveal a reversal of the magnetic field
direction on the arcsecond scale, pointed towards us in the west and
away from us in the east. The electric vectors of linearly polarized
features show an additional trend. Those from the outer parts of
H{\small II} regions implied the direction of the magnetic field
vector closely aligned with the interstellar magnetic field
orientation derived from infrared polarimetry. In contrast, the
vectors measured towards the central parts of H{\small II} regions
were significantly rotated so as to be parallel to the radio jet (region
2). The Zeeman quartet showed a continuing systematic decrease of the
magnetic field strength at a decay rate of
$-0.24\pm0.03$\,per\,cent\,yr$^{-1}$. That is almost half the rate
estimated prevously \citep{cohen90}. However, 
there are also some effects of blending, which cannot easily be removed.    
Observations with
higher angular resolution are needed in order to resolve features blended by
MERLIN.  

\vspace*{0.5cm}
\noindent
{\bf ACKNOWLEDGMENTS} \\
We thank the referee, Prof.\,R.Crutcher, for helpful comments on this work, 
and thank Dr.\,J-M.Torrelles for making available
continuum maps of Cep~A and Prof.\,P.J.\,Diamond for the {\small MFQUV}
files. MERLIN is a national facility operated by the University of Manchester 
at Jodrell Bank on behalf of PPARC. The work was supported by grant 2P03D01122
of the Polish State Committee for Scientific Research.


\label{lastpage}
 
\end{document}